\documentstyle[prb,aps,preprint,epsfig]{revtex}

\begin{document}

\title{Low frequency random telegraphic  noise (RTN) and $1/f$ noise in the rare-earth manganite Pr$_{0.63}$Ca$_{0.37}$MnO$_3$ near the charge-ordering transition}
          
\author{Aveek Bid\footnote[1]{Electronic mail: avik@physics.iisc.ernet.in}, Ayan Guha and A. K. Raychaudhuri\footnote[2]{Electronic mail: arup@physics.iisc.ernet.in}}
\address{Department of Physics, Indian Institute of Science,  Bangalore 560 012,  India.}

\date{\today}
\maketitle

\begin{abstract}
\noindent 
We have studied low frequency resistance fluctuations (noise) in a single crystal of the rare earth perovskite manganite Pr$_{0.63}$Ca$_{0.37}$MnO$_3$ which shows a charge ordering transition  at a temperature $T_{CO}$ $\approx 245K$. The measurements were made across the CO transition covering the temperature range $200K < T < 330K$ and frequency range $10^{-3}Hz < f < 10Hz$. The noise measurements were made using an ac bias with and without a dc bias current imposed on it. We find that the spectral power $S_V(f)$ contains two components - one broad band $1/f$ part that exists for all frequency and temperature ranges and a single frequency Lorentzian of frequency $f_c$  which is strongly temperature dependent. The Lorentzian in $S_V(f)$ which appears due to Random telegraphic noise (RTN) as seen in the time series of the fluctuation, is seen in a very narrow temperature window around $T_{CO}$ where it makes the dominating contribution to the fluctuation. When the applied dc bias is increased beyond a certain threshold current density $J_{th}$, the electrical conduction becomes non-linear and one sees appearance of a significant Lorentzian contribution in the spectral power due to RTN. We explain the appearance of the RTN as due to coexisting Charge ordered (CO) and reverse orbitally ordered (ROO) phases which are in dynamical equilibrium over a mesoscopic length scale ($\approx 30nm$) and the kinetics being controlled by an  activation barrier $E_{a} \approx 0.45eV$. The destabilization of the CO phase to ROO phase causes the nonlinear conductivity as well as the appearance of a RTN type fluctuation when the bias current exceeds a threshold. The $1/f$ noise is low for $T>>T_{CO}$ but increases by nearly two orders in a narrow temperature range as $T_{CO}$ is approached from above and  the probability distribution function(PDF) deviates strongly from  a Gaussian. We explain this behavior as due to approach of charge localization with correlated fluctuators which make the PDF non-Gaussian.
\end{abstract}
\vspace*{0.5cm}
\section{Introduction}
Transport and thermodynamic properties of  colossal magnetoresistive rare earth manganese oxide of ABO$_3$ structure  have attracted considerable current interest. Depending on the composition (Mn$^{4+}$/Mn$^{3+}$), average A-site cationic radius, temperature and magnetic field, the ground state can in general be a ferromagnetic metal (FMM), charge ordered insulator (COI), charge disordered antiferromagnetic insulator (AFI) and even ferromagnetic insulator (FI)~\cite{i1,i2,i3}. In the composition range where Mn$^{4+}$/Mn$^{3+}$ ratio $\approx$ 0.25-0.5, the ground state is typically FMM or COI. In systems like Nd$_{0.5}$Sr$_{0.5}$MnO$_3$ there is a transition from a paramagnetic insulating phase (PI) to a FMM phase which becomes unstable at low temperatures and undergoes a transition towards COI phase. In some narrower band systems like the Pr$_{1-x}$Ca$_x$MnO$_3$ (x $\simeq$ 0.3-0.5), the transition can be from a charge disordered PI phase  to a COI phase. The COI phase so formed can be unstable towards a number of perturbations like magnetic field, current and charge injection~\cite{d1,d2,d3,d4,a3}. The destabilization of the COI state leads to  FMM phase. The COI phase in some narrow band systems like (Nd$_{0.25}$La$_{0.25}$)Ca$_{0.5}$MnO$_3$ or (La$_{5/8-x}$Pr$_{x}$)Ca$_{3/8}$MnO$_3$  ($x\approx 0.35 $) can be unstable at low temperatures and on cooling to T $\simeq$ 100 - 150K gives rise to FMM phase. The brief discussion above shows that in these manganites there are phases with different spin, charge and orbital orders which have almost similar energies~\cite {a1,a2}. Existence of multitudes of phases of similar energies has given rise to the possibility of co-existing phases in certain temperature ranges in these materials. This is generally termed \lq\lq Phase Separation\rq\rq (PS). The PS phenomenon seems to be a rather common phenomenon in  manganites with narrow band-width~\cite{4}. The published reports seem to point to the fact that it can occur over extensive length scales ranging from few nm to  $\mu$m and presumably the microscopic and mesoscopic phase separation have different mechanisms~\cite{5,6,7,8}. Although PS has been seen in a number of systems using both microscopic as well as  bulk techniques it has not yet been established whether it is of electronic origin or it arises from random lattice strains or disorder~\cite{6}. It may also be noted that  often the samples used are of rather poor quality (as shown by the absence of sharp transitions) and for such cases structural disorder driven PS cannot be ruled out. In some of these systems showing PS, there is proximity to a first order transition.

Noise spectroscopy has been used in the past to study the dynamics of phase separation~\cite{n1,n2}. This is a sensitive technique and can give useful information on low frequency fluctuations. In this paper we present a detailed investigation of low frequency (f $\leq$ 10Hz) noise spectroscopy of the Charge Ordered (CO) system Pr$_{0.63}$Ca$_{0.37}$MnO$_3$ which shows a rather well defined transition at T$_{CO} \simeq$ 245K. The principal motivations  for  the present investigation are the following  :

\begin{enumerate}
\item{Extend the noise spectroscopy measurements down to lower frequencies in the $10^{-3}$ Hz range. This is in order to clearly separate out the broadband  $1/f$ component and any narrow band (Lorentzian type) component in the power spectra. Most earlier experiments were carried out for $f> 0.1$ Hz.} 
\item{Doing the experiment near T$_{CO}$ offers certain interesting possibilities. In some CO systems the transition is known to be first order and there is a  clear possibility of co-existing phases at T $\sim $ T$_{CO}$, particularly in the presence of disorder and/or random spins. }
\item{Most of the past investigations of PS using noise spectroscopy were done in CMR compounds where the co-existing  phases are PI (or COI) and FMM or AFI and FMM~\cite{cmr1,cmr2}. In this experiment the measurements are made  near the CO transition where the co-existing phases are PI and COI  or Reverse orbitally ordered (ROO) phases  that occur below $T_{CO}$.}
\item{Application of a dc bias current beyond a certain threshold  leads to non-linear conductivity~\cite{a3,a5}. This arises due to the destabilization of the CO state. This is yet another way of creating co-existing phases. Destabilization of the CO state can also be created by a magnetic field. However, in this experiment we limited ourselves to current induced destabilization. There has been preliminary report of enhanced conductivity  noise  in the regime of non-linear conduction by our group in films of the CO system Nd$_{0.5}$Ca$_{0.5}$MnO$_3$~\cite{n2}. The present studies are more detailed and are done on a more well defined system with a distinct signature of charge ordering in the resistivity ($\rho$) vs T curve. 

We are using a single crystal of  Pr$_{0.63}$Ca$_{0.37}$MnO$_3$ system for our investigations. Pr$_{1-x}$Ca$_{x}$MnO$_3$ shows a charge ordering transition for 0.5$\leq$ x $\leq$ 0.3. However,  the stability of the CO state decreases as  x moves away from half filling (x=0.5).}
\end{enumerate}

We have the following  reasons for choosing the Pr$_{0.63}$Ca$_{0.37}$MnO$_3$ system  :
\begin{enumerate}
\item{The Pr$_{0.63}$Ca$_{0.37}$MnO$_3$ system has been extensively studied by us ~\cite{a3} as well as by other investigators and most of the characteristics of the CO state are known. This is a particularly interesting system because both the trivalent Pr and divalent Ca ions which occupy the A-site  have almost the same ionic radii thus reducing effects of disorder.}
\item{The  single crystal used in this experiment has a particularly sharp transition at T$_{CO}$ and has been used for specific heat experiments by us and the transition has  been shown to be first order where  a substantial part of the entropy of transition appears as latent heat ~\cite{a1}.}
\item{Previous transport studies carried out on this system have shown that at the onset of non-linear conductivity  ferromagnetic filaments appear in the bulk of the solid ~\cite{a5}. Thus there is a clear indication of PS setting in on biasing with a moderate current. In the case of Pr$_{0.63}$Ca$_{0.37}$MnO$_3$ there have also been reports of the co-existence of ferromagnetic and antiferromagnetic fluctuations below T$_{CO}$. Such fluctuations can also lead to PS.} 
\end{enumerate}

As stated before, in our experiment we studied the conductivity fluctuations in presence of an applied dc bias current. Experimentally, we implemented a new feature in order to study noise spectroscopy with an applied dc bias. We have used a setup in which noise can be measured with a fixed amplitude ac signal~\cite{scofield} but one can independently  apply a dc biasing current which does not interfere with the noise measurements. This is a particularly desirable feature. The observed voltage fluctuations need be scaled by the bias that measures the noise in order to obtain the relative fluctuation $S_{V}/V^2$. If the bias for non-linear conductivity and noise measurements are the same, $S_{V}/V^2$ becomes a rather ill-defined quantity. In our set-up (described below) one can  obtain non-linear conductivity and yet retain a well-defined meaning  for $S_{V}/V^2$.

Our investigations show the important result that while there is  broad  $1/f$ component of noise which  exists at all temperatures, there is an additional  low frequency component which becomes very large close to $T_{CO}$. This low frequency component contributes a Lorentzian of corner frequency $f_C$ to the power spectra. This is associated  with appearance of  Random Telegraphic Noise (RTN) in the time domain. $f_C$ is a non-trivial function of the temperature and shows a sharp change near T$_{CO}$. Such low frequency noise components riding on the $1/f$ spectrum also appears when a dc bias is applied  above a threshold current density $J_{th}$. At $J \approx J_{th}$ there is onset of non-linear conductivity in the system.

We also find that as $T\rightarrow T_{CO}$ from above, in the COI state there is a rapid rise in the magnitude of the $1/f$ component of the noise whose magnitude below T$_{CO}$ remains more or less T independent at least down to $T/T_{CO} \approx 0.8$. In the region close to but above the transition where the $1/f$ noise steeply rises as T is lowered we find that the probability distribution function (PDF) changes from a Gaussian to non-Gaussian dependence.

\section{Experimental techniques}
The single crystal used in this experiment has been grown by a floating zone technique. The $\rho$ {\it vs} T curve is shown in figure~1. The charge ordering temperature T$_{CO} \simeq $ 245K. As stated  before the crystal has a first order like phase transition as seen in the specific heat~\cite{a1}  (figure 1 inset). 

The set up used for noise measurements is shown in figure 2. We have used a 5-probe ac arrangement for the measurement of noise using either a transformer preamplifier or a low noise transformer depending on the value of the sample resistance. For sample resistances $<100\Omega$ we used a transformer preamplifier (SR 554) and for  sample resistances $>100\Omega$ we used a low noise preamplifier (SR 560). The carrier frequency was chosen in each case to lie in the eye of the Noise Figure (NF) of the transformer or the preamplifier to minimize the contribution of the transformer noise to the background noise. The output of the preamplifier is fed to a Lock-in-amplifier (SR830). The output low-pass filter of the Lock-in amplifier has been set at 3 msec with a rolloff of  24 dB/octave. For a 3 msec time constant the output filter of the lock-in with 24 dB/octave is flat to $f \leq 10$Hz. This determines the upper limit of our spectral range. The  output of the Lock-in amplifier is sampled by a 16 BIT A/D card  and stored in the computer. At each temperature the data are taken by stabilizing the temperature with $\Delta T/T \simeq 4 \times 10^{-3} \%$. A single set of data are  acquired  typically for a time period of about fifty minutes or more at a sampling rate of 1024 points/sec. The complete data set of  a time series at each temperature consisting of nearly 3 million points  was decimated to about  0.1 million points before the spectral power S$_{V}(f)$ is  determined numerically. The frequency range probed by us ranges from 1 mHz to 10 Hz which allows us to  probe time scales of the order of  15 msec  to  nearly 160 seconds. The frequency range is determined mostly by practical considerations. In addition to the limitation from the output filter of the Lock-in-amplifier, the upper frequency limit  is also  determined by the magnitude of the noise signal coming from the sample with respect to the background (typical background spectral power $\approx$ 4k$_{B}$TR, R is the sample resistance). The lower frequency limit is determined by the quality of the temperature control. In particular in the region of our investigation $dR/dT$ being appreciable, the temperature fluctuation, if not low, can lead to appreciable contribution in the observed voltage fluctuations.  The sample resistance and the bridge output may  show a long time drift. In general such a long time drift is subtracted out by a least square fit to the data. Taking these factors into consideration the lower spectral limit in our experiment has been kept at  $10^{-3}$Hz.

The dc biasing current $J_{dc}$  has been applied using a circuit shown in figure ~2. The decoupling capacitors and inductors shown in the figure decouple the ac and dc source and prevent  dc being applied  to the pre-amplifier circuit. It is important to check that the capacitors and inductor do not affect the gain and the phase of the amplifier and also do not introduce additional features in the power spectra. This was tested by taking data on the sample at room temperature by biasing the circuit using only an ac signal ($J_{dc}$=0) and taking data with the decoupling capacitors and inductors both present and absent. Data were also taken using ac and dc  applied together with the dc bias chosen to stay in the linear region of the E-J curve of the sample. In all these cases power spectra obtained were seen to be identical. In most of the data taken the ac voltage  bias for noise measurement was kept at $\sim$ 1 mV rms. The linear dependence of the observed spectral noise with the ac bias at a typical temperature   is shown in figure ~3 where $fS_{V}$ (measured at f=1 Hz)  is plotted against the rms bias.

 The sample used had dimensions $2mm\times 2mm \times 1mm$. Five gold contact pads were evaporated on the sample and the contacts to the sample were subsequently made by soldering 40 $\mu$m copper wires using Ag-In solder. It may be noted that making low noise and low resistance contacts is always a problem in noise experiments. The best way to check that an appreciable part of the noise is coming from the bulk and not from contact is to establish that the spectral power $S_V(f)/V^2$ scales as 1/$\Omega$ where  $\Omega$  is the volume over which noise is measured. However, with limited sample size, as in the crystal we have, extensive variation of $\Omega$  is not possible. A good check can however be done through monitoring of the background noise. The background noise, when the experiment is carried out in a  well shielded enclosure should have a spectral power close to $S_V \simeq$ 4k$_{B}$TR.   We find that when the contact resistance is low the background noise is close to $\simeq$ 4$k_{B}$TR. However at low temperatures where the sample resistance as well as the contact resistance increases the background noise can become very large. In this condition there is also an appreciable quadrature component of the  signal. We avoided taking data in this region because of the uncertainty that might arise. Ohmic contacts ($I \propto V$) or a quadratic dependence of spectral power $S_V(f)$ on $V$ do not necessarily guarantee the absence of contact noise contribution to the observed noise, especially when the contact resistance is high, as it happens at low temperatures in this case.

\section{Results}
We present the results in three  sub-sections.  In the first sub-section the results for zero applied dc bias are presented ($J_{dc}$ = 0). We then present results of the time domain data. This is followed by the results with an applied dc bias (J$_{dc} \neq$ 0) that induces non-linear conductivity.
\subsection{Noise as a function of temperature and frequency  for $J_{dc}$ = 0}
With no  applied bias the conductivity is ohmic and also the spectral power $S_{V}(f) \propto V^2$ as shown in figure 3. In figure 4(a), we show the spectral power $S_{V}(f)/V^2$ measured at few  representative frequencies as a function of T along with the sample resistivity $\rho$(T). The data are close to the region $220K \leq T \leq 260K$ which is $0.9 \leq T/T_{CO} \leq 1.06$. Note that both $\rho$ and $S_{V}(f)/V^2$ are plotted in log scale. We note that the spectral power at low frequencies ($f \leq 0.125 Hz$) passes through a pair of distinct maxima at the transition temperature ($0.98 \leq T/T_{CO} \leq 1.02$) while  the higher frequency spectral power does not have any such distinct features close to the transition. This observation implies that close to T$_{CO}$, S$_V(f)$ deviates severely from the $1/f $ frequency dependence. We will elaborate on this aspect later on. [Note :  There has been a previous report of electrical noise in Pr$_{2/3}$Ca$_{1/3}$MnO$_3$~\cite{n1}. The experiment carried out at  $f> 0.2$ Hz found the spectra predominantly  $1/f$ type with not much of distinct feature near  T$_{CO}$. This is in agreement with our data for $f \geq 0.2$ Hz.]
In figure~4(b) we show the spectral power as a function of T over an extended temperature scale. It is interesting to see that in this region from $260K < T < 330K$ there is a rapid rise in the magnitude of the noise by more than 2 orders while the spectral power retains its $1/f$ character. 

 In  figure~5 we plot the spectral power as a function of frequency at few representative temperatures ($ 0.94 \leq T/T_{CO} \leq 1.06$). The data are  plotted as $f.S_{V}(f)/V^2$ to accentuate the deviation of the spectral power from the $1/f$ dependence. It can be seen that in this narrow range of temperature,  there is a strong deviation from $1/f$ dependence of the spectral power. The deviation is most visible at $T/T_{CO} \sim 1$. The spectral power regains its predominant $1/f$ character both below and above $T_{CO}$. We find that the spectral power in the region of the CO transition temperature can be fitted by a relation which consists of a 1/f term and a Lorentzian with a corner frequency $f_C$
\begin{equation} 
\frac{S_V(f)}{V^2} = \frac{A}{f} + \frac{B.f_C}{f^2+f_C^2}
\end{equation}
Constants A and B measure the relative strengths of the two terms and are derived from fits to the experimental data. The second term, a Lorentzian, arises from a single frequency fluctuator with a frequency $f_C$. The lines through the data in figure~5 are  fits to equation~1. (The lower limit of $f_C$ is $\approx 2$ mHz as set by our experimental system).  In order to compare the relative strength of the two terms we have obtained the relative resistance fluctuation $<\frac{\delta R^2}{R^2}>$  by integrating S$_V$(f)/V$^2$ within the experimental bandwidth f$_{min}\approx$ 1mHz and f$_{max}\approx$ 10Hz. 
\begin{equation}
<\frac{\delta R^2}{R^2}> = \int^{f_{max}}_{f_{min}} \frac{A}{f}df + \int^{f_{max}}_{f_{min}} \frac{B.f_C}{f^2+f_C^2}df
\end{equation}
\hspace{5.7cm} = $<\frac{\delta R^2}{R^2}>_1 + <\frac{\delta R^2}{R^2}>_2$

The temperature dependence of  the total fluctuation $<\frac{\delta R^2}{R^2}>$,  the contribution of the $1/f$ component $<\frac{\delta R^2}{R^2}>_1$ and that of the Lorentzian $<\frac{\delta R^2}{R^2}>_2$ are shown in figure~6. The figure clearly shows again that the predominant temperature dependence of the noise at $T \sim T_{CO}$ arises from the Lorentzian term while the broadband $1/f $ term is mainly featureless. 

In figure~7 we plot the $f_C$  as  function of $T$ as has been obtained by fitting equation~1 to the data at each temperature. We observe that the corner frequency $f_C$ has a non-trivial temperature dependence.  For $T< 240K$, $f_C$ increases on heating. However, very close to the transition region ($0.98 <T/T_{CO}< 1.02$) the temperature dependence of $f_C$ slows down  with a shallow split   peak at the transition ($T \simeq T_{CO}$). For $T>1.02  T_{CO}$, $f_C$ drops rapidly and $f_{C}\rightarrow$ 0 for $T \geq 260K$ ($T/T_{CO}=1.06$).  f$_C \simeq$ 2 mHz is the limit of our detection on the low frequency side of the spectrum. In the next subsection we present the time series data and we show that the single frequency component is indeed due to RTN. RTN can arise from two level fluctuators (TLF) with two different conductivities in the two levels. In our sample the relative fluctuations in the resistance is $<100 ppm$. This indicates that the two states of the TLF should have comparable resistivities. 

  We associate $f_C$ with the average relaxation rate $\tau ^{-1}_C$ of the TLF fluctuators,  so that $\tau ^{-1}_C$ = 2$\pi f_C$. In this temperature range  the relaxation of TLF is expected to arise   from  thermal activation through a barrier of energy $E_a$.   In that case we can write :
\begin{equation}
f_C = f_0 exp(\frac{-E_a(T)}{k_BT})
\end{equation}
When  the activation energy $E_a$ is constant $f_C$ will increase as $T$ increases. This happens for $T< 240K$. (Above this temperature $f_C$ shows a departure from a simple activated kinetics with  a constant $E_a$.) However, even a small temperature dependence of $E_a$ can change the temperature dependence of $f_C$  since it appears in the exponential. Alternatively, it may also arise from a temperature dependent attempt frequency  $f_0$ although it is a less likely a possibility. The temperature dependent  $E_{a}(T)$ needed to give the observed temperature dependence of $f_C$, obtained from equation~3, is also shown in figure 7. ( The attempt rate $\tau ^{-1}_0$ = 2$\pi f_0 \simeq 6.9 \times 10^{7}sec^{-1}$ was obtained  from the data for $T < 235K$ where $E_a$ is temperature independent and was used as constant for other temperatures.)  It is interesting to note that the temperature dependence of $E_{a}(T)$ is rather shallow for $T<T_{CO}$. It passes through a plateau in the region  $T \sim T_{CO}$ and eventually starts rising  for  $ T > T_{CO}$. The slowing down of the fluctuations close to $ T_{CO}$ can also arise from slowing down expected near a phase transition. The analysis of the experimental data in the time domain shows that at or near the CO transition the spectral power is indeed dominated by a single frequency two level  type fluctuator. In the discussion section we will address the issue whether the existence of the TLF type fluctuator can be associated with PS  near $T_{CO}$ and whether any physical significance can be attributed to $E_a$. 

A plot of the spectral power $S_{V}(f)/V^{2}$ at various frequencies as a function of the sample resistivity is shown in figure 8. In the figure the resistivity $\rho_{CO}$ corresponding to $T_{CO}$ is marked on the resistivity axis. It is seen that the dependence of the spectral power  on the resistivity is very different in the two regions:  $\rho > \rho_{CO}$ ($T < T_{CO}$) and  $\rho < \rho_{CO}$ ($T > T_{CO}$).  In the lower resistance region, $S_{V}(f)/V^2$ is seen to rise sharply with increase in the sample resistivity. It increases by about four orders of magnitude at low frequencies for one order change in resistivity. In this region the spectral power is predominantly $1/f$ type and $S_{V}(f)/V^{2}\propto\rho^{5}$. At $T_{CO}$ , the noise essentially reaches its maximum value and for $\rho > \rho_{CO}$ (below T$_{CO}$) there is a very weak dependence of the spectral power on the  sample resistivity. (Note: The dependence of the spectral power on the resistivity $\rho$ is much stronger than that observed  in usual percolation type transition as has been seen in some of the systems near the COI-FMM transition where the co-existing phases have very different conductances ~\cite{cmr2}. There is no reason that such ideas will be valid for the CO transition  where the co-existing phases, if any, have similar conductances and which is not a percolation type of transition.)

\subsection{Time domain data}
The voltage fluctuation, $\Delta V(t)$, in the time domain  had been recorded as a time series from which the spectral power $S_V(f)$ was obtained. The time series are shown in figure~9 at four  representative temperatures  in the range $ 0.94 \leq T/T_{CO} \leq 1.04$. These are at the same temperatures where f.S$_V$(f)/V$^2$ $vs$ f has been shown in figure~5. It can be clearly seen that at $T \sim T_{CO}$ when the single frequency Lorentzian predominates, we  have the   presence of RTN type jumps in the time series where the voltage fluctuation oscillates  between a \lq\lq high \rq\rq fluctuation  and a \lq\lq low"  fluctuation level. There is the presence of a substantial  $1/f$ component  in the power spectra. As a result the RTN (which has a lower frequency)  modulates the time series of the broadband  $1/f$  noise. Within each \lq\lq high"  and \lq\lq low" state the voltage jump follows the $1/f$ spectra.  The average $\tau ^{-1}_C \simeq$ $\tau^{-1}_{high}$+$\tau^{-1}_{low}$ where $\tau_{high}$ and $\tau_{low}$ are the average times spent in the \lq\lq high" and \lq\lq low" states respectively. Though the $\tau_{high}$ and $\tau_{low}$ are not the same they are very similar in the temperature window studied.  It is clear that the appearance of the single frequency Lorentzian in the spectral power is due to the RTN type behaviour seen in the time domain data in the temperature range close to $T_{CO}$ ($0.96\geq T/T_{CO}\geq$ 1.04). Outside this temperature window it either  does not exist or is not observable due to finite observation time consideration or has very small magnitude beyond the detection limit.

The RTN observed in our sample close to the CO  transition and that observed in $La_{2/3}Ca_{1/3}MnO_{3}$~\cite{cmr1} near the FM-COI transition have both similarities as well as notable differences. Both the samples show RTN and have nontrivial temperature dependence for the relaxation rates for the TLF. However, the relative fluctuation observed for the La system is much larger than that seen in the Pr system. This is explainable because the coexisting phases in the La system being FM and COI phases have different conductivities while in the Pr system the two states of the TLF have similar conductivities both being insulating phases (to be discussed later on). In the Pr system the RTN peaks near the $T_{CO}$ while in the La system it peaks at $T << T_C$. This is presumably due to the fact that the transition in the La-system may have a percolation aspect.

\subsection{Non-linear conductivity and noise}
Preliminary investigation of  non-linear conduction and its relation to electrical noise in manganites has been reported by our group previously in $Nd_{0.5}Ca_{0.5}MnO_{2}$ films~\cite{n2,a6}. Onset of non-linear electrical transport below $T_{CO}$ including a regime showing negative differential resistance have been reported by us before in Pr$_{0.63}$Ca$_{0.37}$MnO$_3$~\cite{a3}. In this report we specifically discuss the question of noise as the electronic transport enters the regime of nonlinear conduction. Our present experimental setup where the dc bias (that controls the non-linear conductivity) and ac bias (that measures the noise) are separated out allows us to do the experiment cleanly.  To our knowledge the noise in these systems have not been  investigated by this technique.

As the bias is increased beyond a threshold value $J_{th}$ the E-J curve begins to deviate from linearity.  This is shown in figure ~10 as an example at T=226K. In this graph we plot the dynamic resistivity $dE/dJ$ as function of the scaled current $J_{dc}/J_{th}$. (For details of non-linear conduction  in Pr$_{0.63}$Ca$_{0.37}$MnO$_3$ as a function of T we refer to our previous work~\cite{a3}.)  We find that for $J_{dc} \approx J_{th}$  a large low frequency component arises in the spectral density similar to that seen close to $T_{CO}$. A typical example of the appearance of excess low frequency  noise in the form of a Lorentzian at the onset of nonlinear conduction is shown in figure~11.  In this figure we have plotted $fS_{V}(f)/V^2$ as a function of $f$  for four  $J_{dc}$ at $T=226K$. We have fitted the power spectra using equation~1. The appearance of a substantial low frequency Lorentzian at $J_{dc} \sim J_{th}$ is apparent. The variation of  $f_C$ with $J_{dc}/J_{th}$ is shown in figure 10. The $f_C$ within the experimental error is essentially unchanged till $J_{dc}/J_{th} \approx 1$. For $J_{dc}/J_{th}> 1$, $f_C$ increases with the applied dc bias.

In the same figure  we have also plotted the relative resistance fluctuations and its two components by integrating the spectral power as in equation~2 at $T= 226K$.  As shown in figure~10, the value of the  $1/f$ component of noise shows no change with applied bias while the excess low frequency component $<\frac{\delta R^2}{R^2}>_2$ shows a jump by nearly  an order of magnitude at $J_{dc}=J_{th}$. We do not take the data at a very high bias because the noise spectra show signs of drifts at higher $J_{dc}$. At the bias used, there is no heating of the sample. This was checked by attaching a thermometer directly on the sample. In the temperature range of the present investigation ($T > 220K$), the detectable heating ($\approx 0.5K$) occurs only for $J_{dc} \geq 1A/cm^2$, which is much larger than the dc bias used by us. (The observed increase in $f_C$ as well as in the relative fluctuation if it would have been from heating would need a heating of $\Delta T \approx 10K$ which is much larger than any measured heating).

The current induced RTN at the onset of non-linear conduction for $T< T_{CO}$ is similar to that which appears for $T\approx T_{CO}$ in zero applied bias. However, the relative fluctuation of resistance is much less in the case of the current induced RTN. The current induced RTN is interesting because it would imply appearance of  the TLF from the coexisting phases whose conversion from one to the other has been kinetically frozen out by the relative large barrier to activation ($E_{a}\approx$ 0.4-0.5 eV). The current presumably destabilizes one of the phases and thus creates a non equilibrium situation. In this context the increase of $f_C$ for $J> J_{th}$ and the onset of non equilibrium condition  may be interpreted  as a suppression of the activation barrier $E_a$  by the applied current. Even a small suppression of $E_a$ is enough to produce the observed shift in $f_C$. However, we have no apriori reason to justify why such a reduction of $E_a$ should happen by the applied current. 

The current induced RTN can be seen only in a narrow temperature window at  $T<T_{CO}$. For $T > 235K$ the $f_C$ of the unbiased sample itself is large. The current causes an upward shift of  the $f_C$ but as $T \rightarrow T_{CO}$ the changes in $f_C$ due to current bias  could not be detected unambiguously. At  much lower T ($T < 200K$)  the $f_C$ is below our spectral detection limit.  However, in this temperature range one can measure the rms noise. In Pr$_{0.63}$Ca$_{0.37}$MnO$_3$ for $T < 150K$ at current densities $J_{dc} >> J_{th}$ there is onset of a region of negative differential resistance. In figure~12 we plot the $E$ vs $J$ curve along with the rms fluctuation as a function of  $J$. The fluctuation peaks near the $J_{th}$  and becomes large again at the region of negative differential resistance. In the same figure we show the same data taken on a charge ordered film of Nd$_{0.5}$Ca$_{0.5}$MnO$_{3}$ for comparison~\cite{n2}. The onset of large fluctuations at the threshold current density for  non-linear conduction is thus a general phenomena. (The film of  Nd$_{0.5}$Ca$_{0.5}$MnO$_{3}$, like other films of CO materials, may not have long range charge ordering unlike Pr$_{0.63}$Ca$_{0.37}$MnO$_{3}$. That may explain the low value of $J_{th}$ in them.)

Before we end this section we summarize the important results:
\begin{enumerate}
\item{There is appearance of large low frequency fluctuations near $T_{CO}$. The low frequency noise has characteristics of  Lorentzian  at a single frequency $f_C$. $f_C$ have a non-trivial T dependence. The time domain data can relate this Lorentzian in power spectra as arising from RTN.}
\item{There is coexistence of the RTN  and $1/f$ type fluctuations and they have distinctly different T dependences. It appears that they may arise from different origin  and different length scales.}
\item{Application of $J_{dc}$ leads to onset of non-linear conductivity. The onset of non-linear conductivity also leads to a  low frequency RTN type noise as seen close to $T_{CO}$ in zero bias.} 
\item{The noise above $T_{CO}$ which has a $1/f$ spectral dependence increases rapidly as the CO transition is approached from above. However, the $1/f$ noise has a shallow temperature dependence below $T_{CO}$. }
\end{enumerate}

\section{Discussion}
We note that the spectral power has two distinct components - the $1/f$ component (present at all T) and the Lorentzian component arising from RTN type fluctuation occuring within a narrow temperature window around $T_{CO}$. From the observed data it is a reasonable assumption that these two arise from two different sets of fluctuators making independent contributions. In the discussion below we would like to probe the origin of these two sets of fluctuators which presumably may occur at different length scales.

The appearance of  RTN type fluctuation  can be associated with co-existing phases of different conductivity. If the two coexisting phases contribute to the conductivity fluctuation in the time scale of our experiment, there should be dynamics associated with these phases in this time scale. This  slow dynamics can come from a transformation from one phase to the other. This transformation will take place through an energy barrier by thermal activation. This is the general picture that is used to justify existence of low frequency noise in the region with phase separation. The important question is what are the coexisting phases at the CO transition? In CMR systems in which the noise has been investigated the coexisting phases can be FM and CO phases~\cite{cmr2}. But the FM phase does not appear here. We seek clue to this question from two structural studies using neutron~\cite{6} done on the CO system Pr$_{0.7}$Ca$_{0.3}$MnO$_{3}$ and TEM~\cite{pod1}done on a related system La$_{0.225}$Pr$_{0.4}$Ca$_{0.375}$MnO$_{3}$ which shows a transition from CO to FM phase.

The neutron study~\cite{6} on Pr$_{0.7}$Ca$_{0.3}$MnO$_{3}$ found that below $T_{CO}$ two phases that coexist are the charge ordered phase with $e_g$ orbitals  oriented in the a-c plane and the reverse orbitally ordered (ROO) phase with the $e_g$ orbitals oriented perpendicular to the a-c plane. The two phases have opposite orthorhombicity and can coexist due to strain consideration where the strain created by the CO phase is balanced by the opposite strain of the ROO phase. This can build up a strain mediated self organized structure~\cite{little1}. The nucleation of the ROO phase occurs just below $T_{CO}$ and these two phases coexist in almost equal proportion. The coexistence of these two phases occur in a mesoscopic scale 50-200 nm. Both phases are charge localized although the ROO phase has less degree of charge localization and is expected to have a larger conductivity. 

TEM studies~\cite{pod1} of  the La$_{0.225}$Pr$_{0.4}$Ca$_{0.375}$MnO$_{3}$ system showed that in addition to the FM and COI phases there are coexisting charge disordered insulating phase (CDI). The large RTN  (in the time scale of few seconds) near the FM-COI transition was found to arise from the coexisting COI-CDI phases. The CDI phase occurs in a length scale of 10-20 nm. 

In view of the above we propose that the RTN observed by us in Pr$_{0.63}$Ca$_{0.37}$MnO$_{3}$ close to the $T_{CO}$ arises from coexistence of CO and ROO phases. Both these phases are charge localized and are both insulating. However, the degree of charge localization is less in the ROO phase which has more degree of disorder~\cite{6}. One would expect the ROO phase to have higher conductivity although the two phases will have similar order of conductivity.  The relatively low magnitude of the RTN ($<100ppm$) can thus be explained from the similar conductivities of the two states of the TLF.

The RTN needs cooperative switching between the two states and this would need transition between coexisting states over mesoscopic length scales. Such coexistence  can come from strain accommodation between the two phases with opposite volume strain which can self organize to a mesoscopic scale~\cite{little1}. The RTN would arise form transformation between the two phases which are kinetically stabilized by an energy barrier. The observed activation energy $E_a$ associated with the $f_C$ can thus be interpreted as the energy barrier between the two phases. We can obtain an estimate of the length scale ($L_c$) over which the transformation occurs  from the value of $E_a$($\approx 0.45 eV$). From the neutron studies of the two co-existing phases~\cite{6}, we find that the transformation would involve a strain accommodation of the order of $\epsilon \approx 10^{-3}$. The bulk modulus of these oxides are in the range of $\approx 100-200$ GPa. This gives an energy density $\approx$ 50-100 kJ/m$^3$ associated with the transformation. Assuming a spherical transformation volume of diameter $L_c$ and an activation energy of $E_{a}\approx$ 0.45 eV we obtain an estimate $L_{c} \approx 30nm$.  (A somewhat larger estimate of $L_{c} \approx 40-50 nm$ will be obtained if the strain is accommodated through a shear process, the shear modulus being much smaller $\approx$ 30GPa.) This value agrees very nicely with the scale of mesoscopic phase separation seen in these solids. The small variation in $E_a$ that would be needed to explain the T dependence of  $f_C$ can arise even if there is a small increase in the accommodation strain between the two phases as $T \rightarrow T_{CO}$. 

The observed attempt frequency $f_0$ is rather low ($\approx 10^7 sec^{-1}$). This is much less than the typical magnitude of the jump frequency ($10^{12}-10^{14} sec^{-1}$) seen in relaxation processes that involve one or few atoms. The low value of $f_0$ or the relaxation rate $\tau_{0}^{-1}$ suggests that the process involves mesoscopic length scales. The rearrangement process would involve propagation of the strain field  created by the transformation process over a length scale $L_c$. This would determine the scale of jump frequency $f_0$. (Alternately this may also mean that there may be an entropy term that must also be considered in addition to the activation energy term.)

The coexistence of the CO and ROO phases can also  justify the observation of current induced RTN. At any given temperature there is an equilibrium volume fraction of the two phases. Below $T_{CO}$ the transformation between them is kinetically frozen.  The current induced RTN accompanying non-linear transport will imply that on application of the $J_{dc}$ one of the states become unstable and this changes the equilibrium concentration of the two phases. In this context we recall that in La$_{0.225}$Pr$_{0.4}$Ca$_{0.375}$MnO$_{3}$ it has been observed that~\cite{pod1} on exposure to electron current in an electron microscope the CO phase breaks down to CDI phase of length scales $\approx 20nm$ once the current exceeds a threshold of $\approx$ 0.5A/cm$^2$. This is what is expected also in the coexisting phases under a bias current. It is likely that the CO phase breaks down to more conducting ROO phase on current bias thus reducing its resistivity and enhancing the RTN. Interestingly the threshold current level seen by us $J_{th}\approx$ 0.1-0.5 A/cm$^{2}$ for nonlinear conduction  matches very well with the results of electron induced CO instability~\cite{pod1}. 

The strain induced self organization is a feature seen in alloys showing shape memory. We have recently studied the resistance noise in the shape memory alloy NiTi as it is taken through the Austanite-Martensite transformation~\cite{niti}. Interestingly we find that at the transition the noise becomes very large and the spectral power shows large deviation from the simple $1/f$ power dependence at low frequencies ($f \leq 0.1 Hz$). The large deviation of the observed $S_V(f)$ in NiTi near the Austanite-Martensite transformation thus has a similarity with the spectral power seen near the $T_{CO}$ in our sample.This issue is currently under investigation.
 
The existence of $1/f$ noise does not need existence of coexisting phases at mesoscopic scale and it can appear from processes that occur at the atomic level. In our sample $1/f$ noise occurs over the whole temperature range, however, the most notable feature is the rapid increase of $1/f$ noise as the $T_{CO}$ is approached from above. Increase in $1/f$ noise in insulators on cooling is generally seen as a reflection of the reduction in the carrier density on cooling. In that case one would expect a direct correspondence of the $S_V(f)$ and $\rho$. In our sample however, we find a virtual decoupling of the two while the noise increases much faster than $\rho$ as $T \rightarrow T_{CO}$ from above and it is essentially temperature independent below $T_{CO}$ where $\rho$ continues to rise following an activated behavior.

We made another very interesting observation in the probability distribution function (PDF) of the voltage fluctuation. We find that while at high temperature the PDF can be described by a Gaussian, as $T_{CO}$ is approached the PDF shows non-Gaussian tail eventually taking the form of a Poissonian just above $T_{CO}$ ($T/T_{CO} \sim 1.04$). In the region close to $T_{CO}$ ($T/T_{CO} \leq 1.04$) the PDF is dominated by the RTN. However, below $230K$ when the RTN contribution is less again the PDF of the $1/f$ noise  can be studied and we find that it has regained its Gaussian form. The decoupling of the $1/f$ noise from the resistivity and appearance of non-Gaussian fluctuation points to building up of a correlation in the charge disordered liquid as the carrier  is getting frozen in the CO state. We propose the following scenario for the unusual temperature dependence of the $1/f$ noise although we do not have an  independent supporting evidence for this scenario. The localization of the charges or groups of charges around specific lattice sites can lead to a local Coulomb potential that can relax over long time scales. This can lead to a low frequency charge dynamics that can enhance the low frequency noise. As the sample  is getting cooled and  the charge is getting localized, the relaxation time of the charge fluctuation will increase leading to enhanced low frequency spectral power. The freezing of the charges in the CO state in this case will be a glass like freezing of the charge liquid.  Below $T_{CO}$ the frozen-in defects in the charge lattice will provide the path to relaxation just like it happens in atomic relaxation in a structural glass and this will  make $S_V(f)$ temperature independent. This behavior is under detailed investigation and we refrain from making further comments on this. 

To conclude, we have carried out comprehensive studies on low frequency conductance noise in a CO system near the CO transition temperature. We find that there are two types of noise (RTN and $1/f$) with distinct characteristics that accompany the CO transition. The RTN occuring close to CO transition or occuring due to current induced destabilization of the CO state is due to coexisting CO and ROO phases. This is a mesoscopic phenomena occuring over an estimated length scale of $\approx 30nm$. The $1/f$ noise most likely occurs at a microscopic length scale and it is strongly affected by charge localization at the onset of CO transition.

One of us (AKR) wants to acknowledge DST, Government of India for a sponsored project. We would like to acknowledge stimulating discussions with Prof. CNR Rao on phase separation and also for kindly providing us with the sample.

\newpage

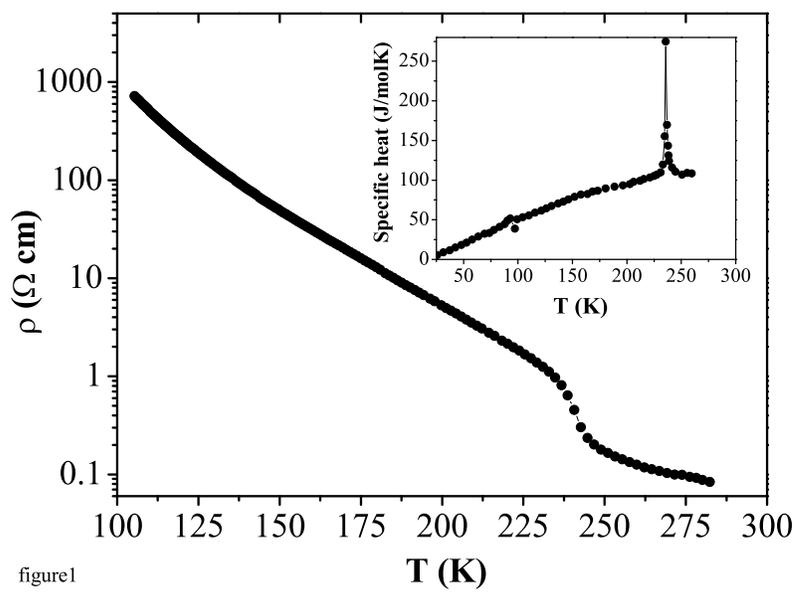
figure ~1. The charge ordering transition in Pr$_{0.63}$Ca$_{0.37}$MnO$_3$  as seen by the resistivity and  specific heat.

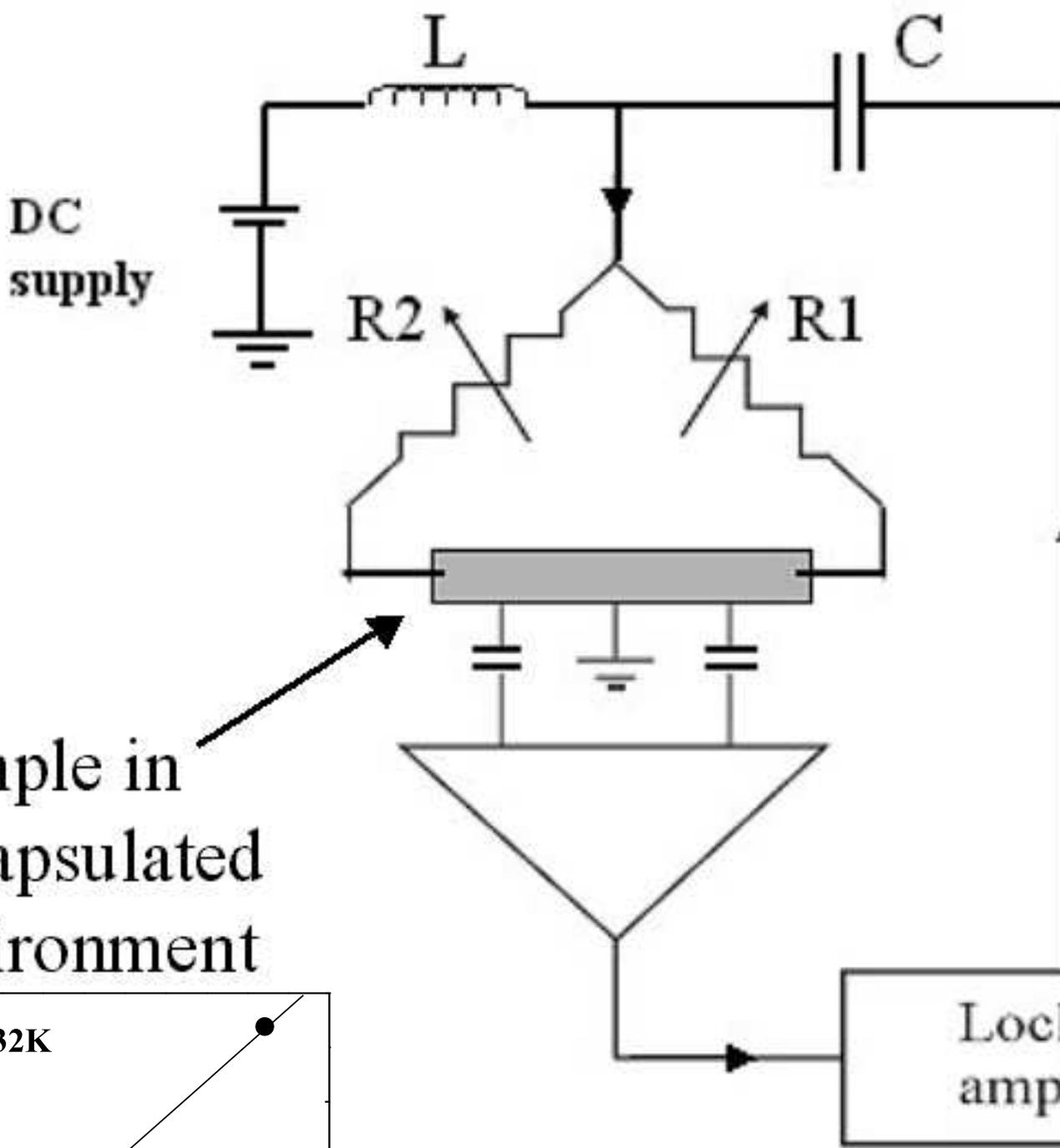
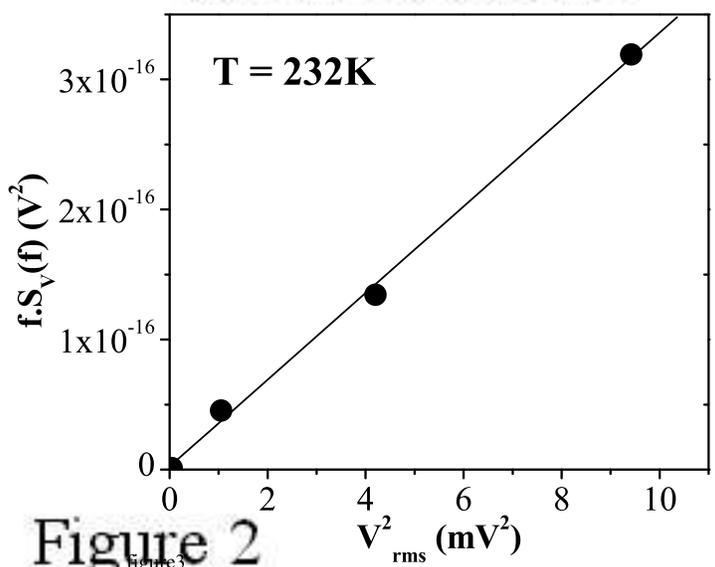
figure ~2. The experimental setup that allows measurement of the noise by an ac bias with  a superimposed dc. 

figure ~3. $fS_{V}(f)$ as function of the ac bias.  

figure ~4(a). The spectral power $S_{V}(f)/V^2$ measured at few  representative frequencies as a function of T along with the sample resistivity $\rho$(T) close to the CO transition, figure ~4(b) the same over a wider temperature range.

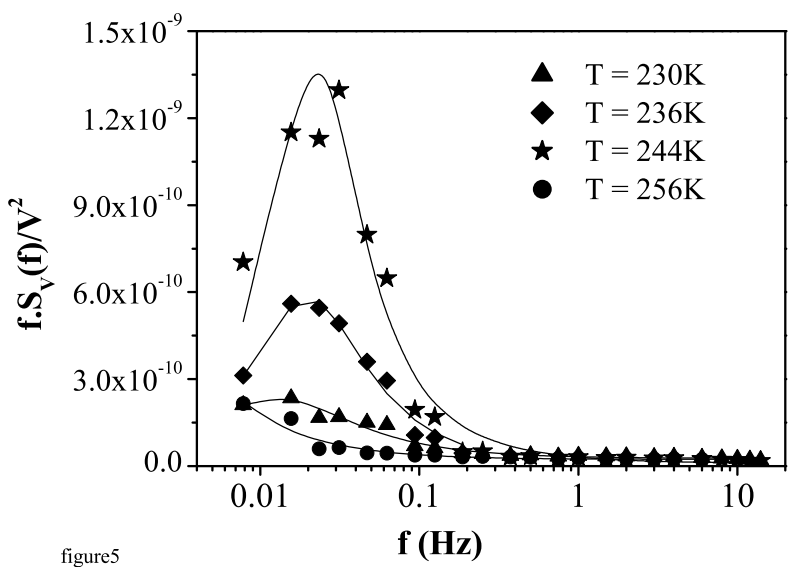
figure~5. The spectral power as a function of frequency at few representative temperatures ($ 0.94 \leq T/T_{CO} \leq 1.04$)
The appearance of a distinct Lorentzian riding on the $1/f$ background is seen. The lines are fit to eqn~1.

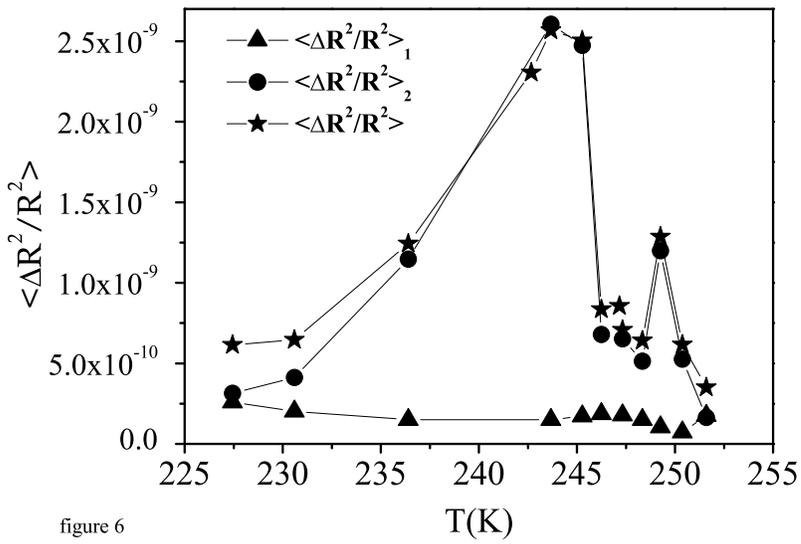
figure~6.  The temperature dependence of  the total fluctuation $<\frac{\delta R^2}{R^2}>$,  the contribution of the $1/f$ component $<\frac{\delta R^2}{R^2}>_1$ and that of the Lorentzian $<\frac{\delta R^2}{R^2}>_2$ (see text and eqn~2)

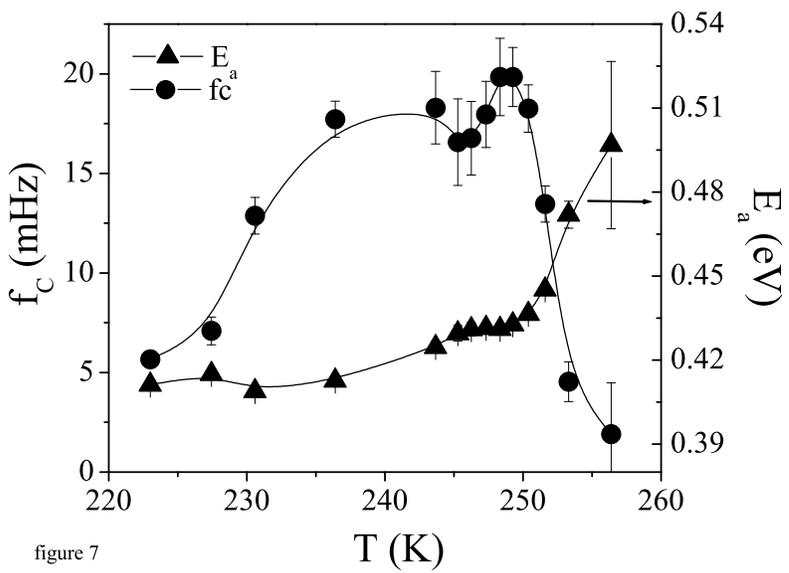
figure~7.  $f_C$  as  function of $T$. The $E_{a}(T)$ needed to give the observed temperature dependence of $f_C$, obtained from equation~3, is also shown.  

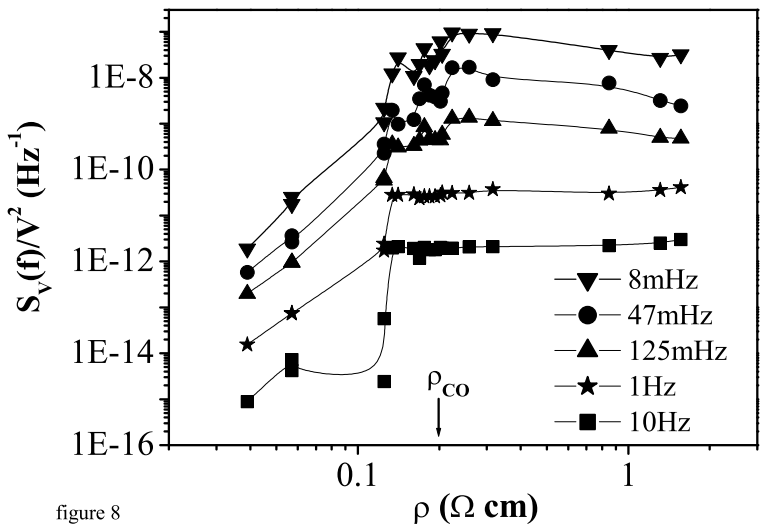
figure~8. Conductance fluctuations at various frequencies as a function of the sample resistivity . The resistivity $\rho_{CO}$ corresponding to $T_{CO}$ is marked on the resistivity axis.

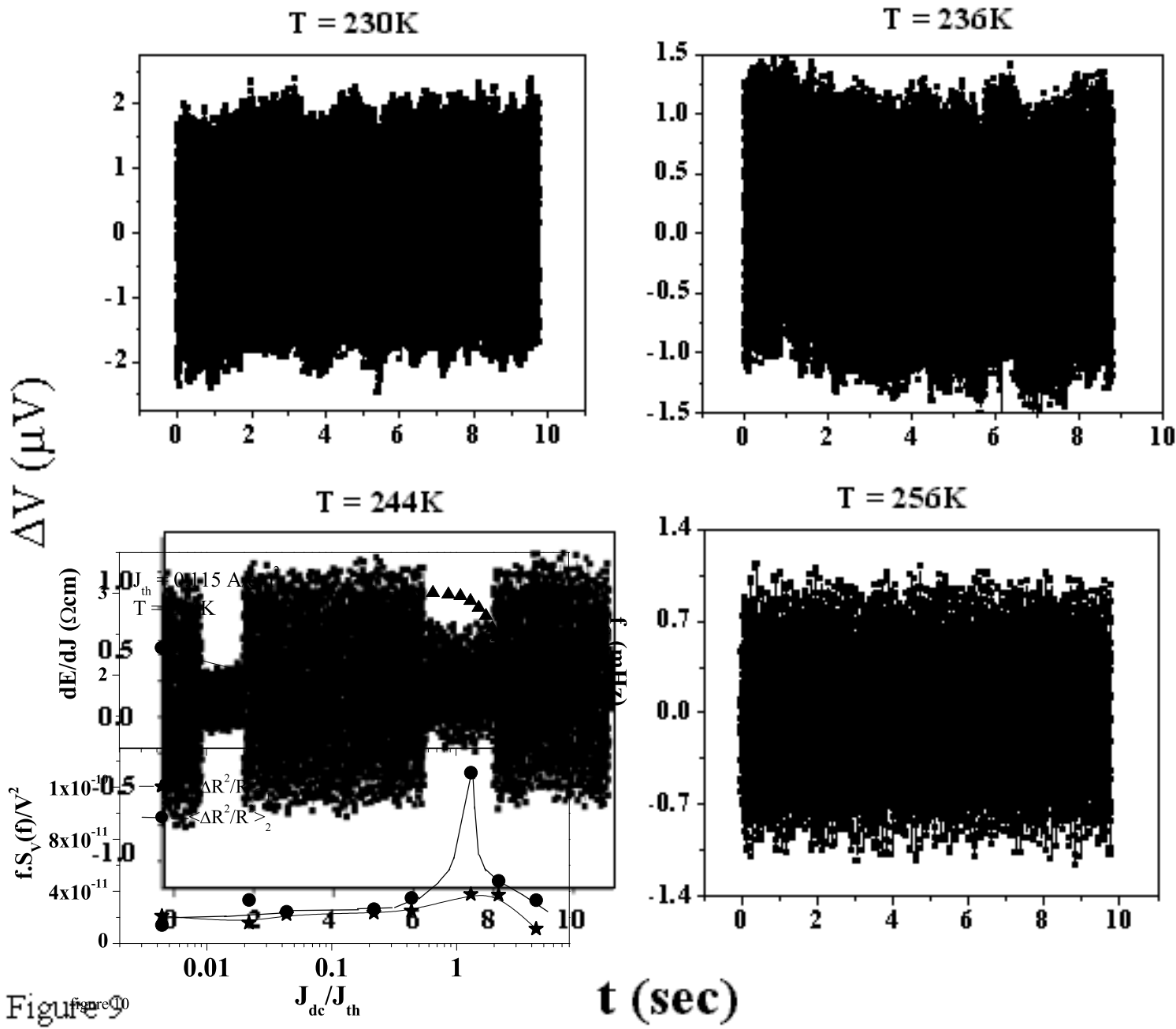
figure~9. The time series  at four  representative temperatures  in the range $ 0.94 \leq T/T_{CO} \leq 1.04$

figure~10. Dependence of the dynamic resistivity $dE/dJ$, $f_C$ and the  contribution of the $1/f$ component $<\frac{\delta R^2}{R^2}>_1$ and that of the Lorentzian $<\frac{\delta R^2}{R^2}>_2$ as function of the dc bias $J_{dc}$ at T=226K.

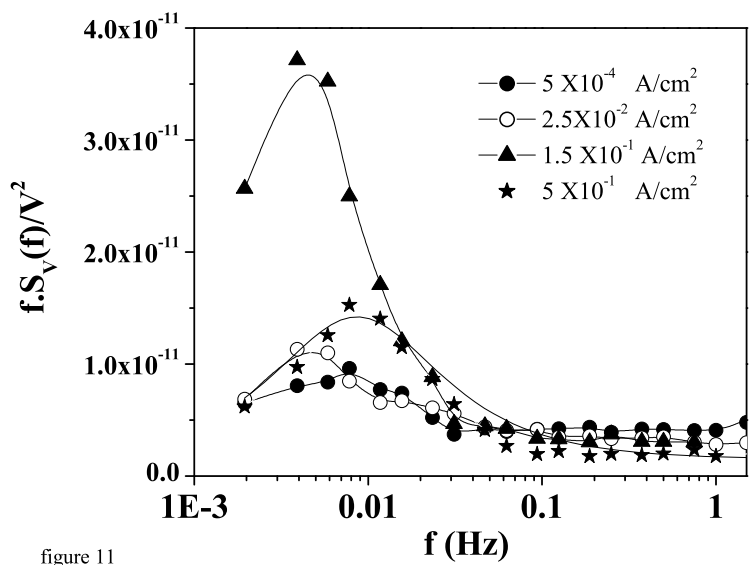
figure~11. A typical example of the appearance of excess low frequency Lorentzian  noise  at the onset of nonlinear conduction at T=226K. The lines are fit to eqn~1.

figure~12. $E$ vs $J$ curves along with the rms fluctuation as a function of  $J$. The fluctuation peaks near the $J_{th}$  and becomes large again at the region of negative differential resistance. In the same figure we show the same data taken on a charge ordered film of  Nd$_{0.5}$Ca$_{0.5}$Mn O$_{3}$ for comparison.

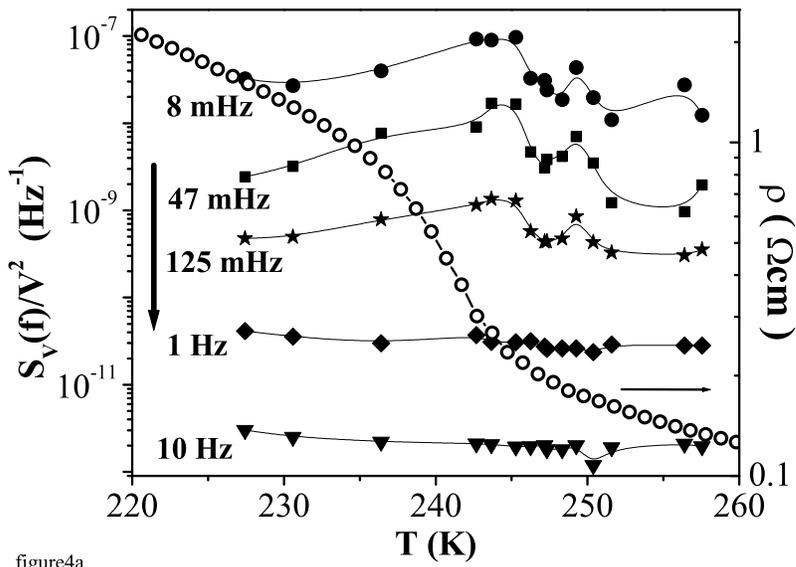
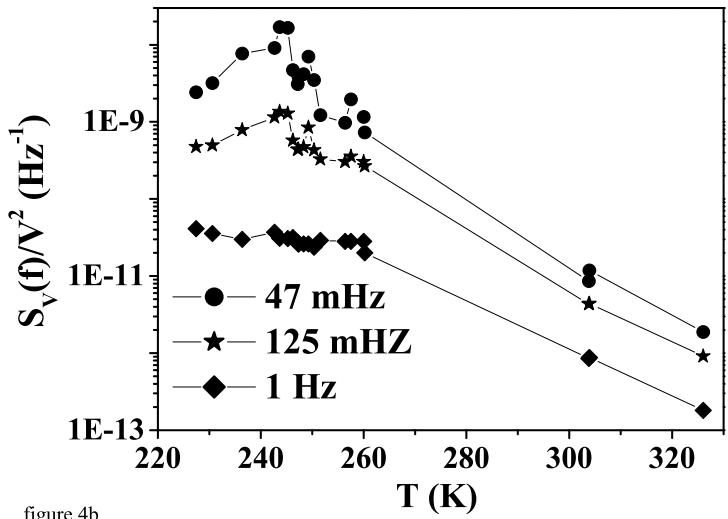
\end{document}